\documentclass{article}

\usepackage{amsmath}  
\usepackage{amsfonts} 
\usepackage{amssymb}  
\usepackage[normalem]{ulem}

\usepackage[final]{neurips_2024}

\usepackage{fancyhdr}
\fancypagestyle{urls}{
  \fancyhf{} 
  \fancyfoot[C]{\thepage} 
}

\usepackage{natbib}
\usepackage[utf8]{inputenc} 
\usepackage[T1]{fontenc}    
\usepackage{hyperref}       
\usepackage{url}            
\usepackage{booktabs}       
\usepackage{amsfonts}       
\usepackage{nicefrac}       
\usepackage{microtype}      
\usepackage{xcolor}         
\usepackage{pgfplots}
\usepackage{wrapfig}
\usepackage{pgf}
\usepackage{booktabs}
\usepackage{pgfplots}
\usepackage{pgfplotstable}
\pgfplotsset{compat=1.17}
\usepackage{graphicx}
\usepackage{fancyvrb}
\usepackage{fancyhdr}
\usepackage{hyperref} 

\usepackage{lipsum} 

\fancypagestyle{code}{
  \fancyhf{} 
  \fancyfoot[L]{\footnotesize{*Code at https://github.com/himansh005/persona\_lora}} 
}

\usepackage{pgfplots}
\usepackage{listings}
\usepackage{xcolor}

\usepackage{tcolorbox}
\tcbuselibrary{listings, breakable}

\newtcblisting{customlisting}[1][]{
    breakable,
    listing only,
    colback=lightgray!20,
    colframe=black,
    listing options={
        basicstyle=\ttfamily\small,
        breaklines=true,
        breakatwhitespace=false,
        columns=fullflexible
    },
    title=#1
}
\title{Personas within Parameters: Fine-Tuning Small Language Models with Low-Rank Adapters to Mimic User Behaviors}

\author{
  Himanshu Thakur \\
  JPMorgan Chase \& Co\\
  Palo Alto, CA 94304 \\
  \texttt{himanshu.thakur@chase.com}
  \And
  Eshani Agrawal \\
  JPMorgan Chase \& Co\\
  Palo Alto, CA 94304 \\
  \texttt{eshani.agrawal@jpmchase.com}
  \And
  Smruthi Mukund \\
  JPMorgan Chase \& Co\\
  Palo Alto, CA 94304 \\
  \texttt{smruthi.mukund@chase.com}
}

\begin{document}

\maketitle

\begin{abstract}

A long-standing challenge in developing accurate recommendation models is simulating user behavior, mainly due to the complex and stochastic nature of user interactions. Towards this, one promising line of work has been the use of Large Language Models (LLMs) for simulating user behavior. However, aligning these general-purpose large pre-trained models with user preferences necessitates: (i) effectively and continously parsing large-scale tabular user-item interaction data, (ii) overcoming pre-training-induced inductive biases to accurately learn user specific knowledge, and (iii) achieving the former two at scale for millions of users. While most previous works have focused on complex methods to prompt an LLM or fine-tune it on tabular interaction datasets, our approach shifts the focus to extracting robust textual user representations using a frozen LLM and simulating cost-effective, resource-efficient user agents powered by fine-tuned Small Language Models (SLMs). Further, we showcase a method for training multiple low-rank adapters for groups of users or \textit{persona}, striking an optimal balance between scalability and performance of user behavior agents. Our experiments provide compelling empirical evidence of the efficacy of our methods, demonstrating that user agents developed using our approach have the potential to bridge the gap between offline metrics and real-world performance of recommender systems.

\end{abstract}

\section{Introduction}
In the rapidly evolving landscape of recommendation systems, user behavior simulation has become crucial for cost, safety, and efficiency benefits. A promising line of work towards this has been Agentic-LLM systems to simulate a user in a recommendation environment. While prior work has been able to effectively establish the value of such systems, representing a user faithfully and at scale has for a large part been unsolved. Thus, we posit that developing robust user simulation agents requires the ability to continuously capture a user's static and dynamic behavior from large volumes of noisy interactions.  

To address these challenges, we propose a method to build scalable and efficient user behavior agents. First, we transform large volumes of user interactions into meaningful textual representations. Next, we use low-rank adapters to fine-tune a SLM on this data, creating a user agent that closely mimics actual user actions. While previous works have employed Retrieval Augmented Generation (RAG) systems to build agents, we argue that fine-tuning an SLM is more effective and scalable in real-world applications. Additionally, we show that users can be grouped into personas, balancing the number of parameters required with the personalization quality of these user agents. Thus, our main contributions can be summarized as follows:

1. We demonstrate a hierarchical knowledge distillation process that utilizes an LLM and self-reflection to convert tabular user interactions into textual user profiles and ground their interactions in reason. We find that both of these components are key to improving the user agent's performance.\\
2. We show that low-rank adaptation of SLMs can match or exceed the performance of frozen LLMs for building personalized agents. Furthermore, we show that Retrieval Augmented Fine-tuning (RAFT) helps in the effective utilization of short-term and long-term memory needed to build the user agent.\\
3. Finally, we show that clustering users based on their profiles generated during distillation helps strike a balance between personalization performance and the number of model parameters needed. 

\begin{figure*}[t]
    \centering
    \includegraphics[width=1\textwidth]{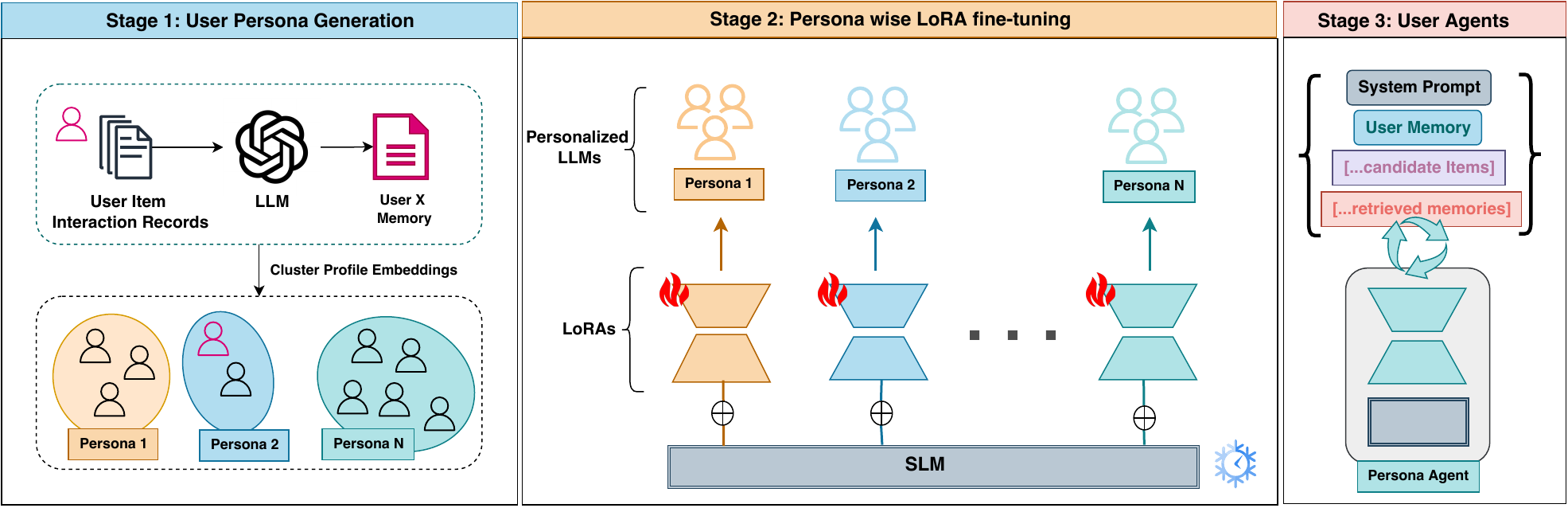}
    \vspace{-0.5cm}
    \caption{An overview of our proposed approach. First, we iterate over all user-item interaction records to generate user profile features and explanations for certain unique interactions. Then, we cluster users based on their profile embedding to generate N personas. In stage 2, we train a low-rank adapter on all user interaction data within the persona keeping the base SLM weights frozen. Finally, in stage 3, we utilize at persona-level SLMs to build user agents.}
    \label{fig:mainfig}
    \vspace{-0.2cm}
\end{figure*}

\thispagestyle{code}
\vspace{-1em}
\section{Related Works}

In recent years, the field of LLMs has witnessed significant advancements, particularly in the domains of personalization, simulation of human behavior, and recommendation systems. This section reviews the most pertinent works, categorized by their thematic focus. 

\textbf{Simulating User Agents using LLMs:} The simulation of human behavior through LLMs shows promise for interactive systems. Jiang et al. \cite{jiang2023personallm} and Park et al. \cite{park2023generative} enhance virtual agents with personality traits and lifelike interactions. Wang et al. \cite{wang2023recmind} and Yang et al. \cite{yang2024generate} improve recommendation accuracy and adaptability, highlighting LLMs' potential in enhancing user interactions and recommendations. Our works aims to progress this research for SLMs.

\textbf{Enriching Users and Items using LLMs:} Sun et al. \cite{sun2024persona} propose a framework to enhance user representation by distilling interactions into a hierarchy of user descriptions. This method leverages persona-specific data to improve the contextual appropriateness of response predictions. Zhang et al. \cite{zhang2024agentcf} explore the use of generative agents in recommendation systems, treating items as agents that engage in multi-turn interactions with users, thereby enriching each other's metadata. This approach proves particularly effective in cold start scenarios and demonstrates robust generalization capabilities. We employ a combination of both approaches in this work to generate enriched user and item information.

\textbf{Parameter-Efficient Tuning for Personalization:} Tan et al. \cite{tan2024democratizing} use low-rank adapters to create instruction-tuned LLMs for each user, enhancing accuracy and generalization of user behavior prediction. Our approach builds on this by incorporating comprehensive user profiles and enriched interactions to better encode preferences. Also, our aim is to check the effectiveness of the appraoch on SLMs. The minimal updated parameters and plug-and-play nature of PEFT make it ideal for efficient SLM personalization. Our work embeds user history within user level LoRA parameters, providing each user agent with a robust and expressive long-term memory.

\textbf{LLMs for Personalization:} Current research on LLM personalization focuses on crafting prompts using historical user data to customize responses \cite{chen2023personalization}. Strategies include: (1) Vanilla Personalized Prompts: Use in-context and few-shot learning to embed user behavior history (e.g., personal rating histories by Dai et al. \cite{dai2023chatgpt} and Kang et al. \cite{kang2023user}, task-specific interaction histories by Liu et al. \cite{liu2023chatgpt}).
(2) Retrieval-Augmented Personalized Prompts: Retrieve relevant user history to enhance prompts (e.g., LaMP \cite{salemi2023lamp}, AuthorPred \cite{li2023teach}, Pearl \cite{mysore2023pearl}).
(3) Profile-Augmented Personalized Prompts: Summarize user preferences into natural language profiles (e.g., Richardson et al. \cite{richardson2023integrating}, ONCE \cite{liu2023once}).
These methods aim to improve the personalization of LLM responses. While these advancements are notable, the current methods struggle to effectively leverage long-term user preferences especially in SLMs. Our approach overcomes this limitation by embedding long-term preferences directly into the model parameters using the personal PEFT module. 

\section{Methodology}

In this section, we define the standard notation and methodologies used throughout the paper. We begin by introducing the fundamental concepts and terminologies that form the basis of our approach. This includes the representation of users, items, and their interactions within a traditional recommender system. We then describe our approach on construction and training of user agents to predict user preferences.
\vspace{-0.5em}
\subsection{Preliminaries}
\vspace{-0.5em}
In a traditional recommender system, we have a set of users \( U = \{u\} \), a set of items \( I = \{i\} \), and a set of their interaction records denoted by \( D = \{\langle u, i \rangle\} \). In our system, a user agent is built using an open source SLM and has short-term and long-term memories, denoted by \( M_s \) and \( M_l \), respectively. Thus, a user agent can be defined as:
\vspace{-0.3em}
\[
y_i \leftarrow f_{\text{SLM}}(M_s, \text{item}_i, \{\text{item}_1, \ldots, \text{item}_k\}),
\]
\vspace{-0.2em}
where \( M_s \) is the user agent's short-term memory, \(\text{item}_i\) is the description of the \(i\)-th item, and \(\{\text{item}_1, \ldots, \text{item}_k\} \subseteq M_l\) are items retrieved using \(\text{item}_i\) as a query by a similarity function \( m(e(\text{item}_i), e(\text{item}_k)) < \delta \). We implement \( m \) as cosine similarity and \( e \) using a frozen embedding model, with \( \delta \) being a threshold determined heuristically. Finally, \( y_i \) is the response of the SLM-based user agent and can have multiple tokens in its output.
\vspace{-0.5em}

\subsection{Distilling User Preferences and Enriching Interactions}
\vspace{-0.5em}

A key challenge of using SLMs for building user agents has been the appropriate representation of the user. Inspired by \cite{sun2024persona} and \cite{zhang2024agentcf}, we use an hierarchical self-reflection based algorithm that inputs a chronologically sorted user-item interaction dataset in batches \( D \) and outputs \( D_{\text{enriched}} = \{\langle u_{\text{enriched}}, i_{\text{enriched}} \rangle\} \), where the user and item descriptions reflect the preferences of the user and the item, respectively. The process results in generation of these two key features per user:\\
\\
\textbf{User Profile (\( M_s \))}: A description of general traits of the user inferred based on their interaction history. This act as the user's short term memory \( M_s \).\\
\\
\textbf{Enriched User Interaction (\( M_l \))}: A textual description that explains the why the user liked or disliked a given item. 
To create a persona agent with an SLM, we use \( M_s \) as a fixed system prompt during training and inference. We also use user profile embeddings to identify user clusters or personas. Additionally, \( M_l \) is embedded and stored in a database for use in RAG and RAFT. During training, we fetch the \( k \) nearest items using the current item embedding and use these as queries to retrieve their enriched versions.
\vspace{-0.5em}

\subsection{Personalized Low-Rank Adapters}
\vspace{-0.5em}

Central to our work is parameter efficiency, as we aim to simulate a very large number of users, which is typical in a recommender system. Previous research \cite{tan2024democratizing} has demonstrated that training multiple low-rank adapters per user can effectively capture user-specific knowledge and style. However, the number of parameters required for this approach would be enormous. Therefore, we propose that multiple users can be clustered into specific behavioral groups or personas, and training a low-rank adapter for each group could suffice.

We group users into \( K \) (note the difference in notation for \( k \) retrieval items) personas and fine-tune an SLM using K seperate low-rank adapters, one per persona, on a downstream user-item interaction task. Once trained, we load the LoRA corresponding to the user's persona during inference to predict the target(s). We study the effects of these two types of memory with two fine-tuning variants: 

\textbf{Using \( M_s \) only}: For a user \( U \), we train a low-rank adapter on the enriched user persona from the distillation step given as: \( y \leftarrow F_{\theta_{u_k}}(\texttt{user\_profile}, \texttt{movie\_to\_rate}) \). 

\textbf{Using \( M_s \) and \( M_l \)}: For a user \( U \), we train a low-rank adapter on the enriched user profile and top-\(k\) relevant item interactions retrieved by a similarity function \( m(e(U+I), e(I+I')) < \delta \) given as: \( y \leftarrow F_{\theta_{u_k}}(\texttt{user\_profile}, \texttt{movie\_to\_rate}, \texttt{top\_\( k \)\_enriched\_interactions}) \). We also perform experiments with training a single LoRA for all users, called \(\theta_{\text{single}}\), to compare with our method of training a separate LoRA for each persona, called \(\theta_{\text{persona}}\).



\section{Experiments and Results}

In this section, we detail the models, datasets, and methodologies used to evaluate the effectiveness of our proposed approach. For distillation and persona generation, we utilize GPT-4o to generate \( M_s \) and \( M_l \), and text\_ada\_002 embeddings to embed \( M_s \) and \( M_l \). We employ LLaMA-3-8B-Instruct (referred as \texttt{LLaMA-3}) as our LLM and Phi-3-Mini-4k-Instruct as our SLM (referred as \texttt{Phi-3-Mini}), which has 3.8 billion parameters and a context length of 4,092 tokens. To evaluate our approach, we use the MovieLens-1M dataset \cite{10.1145/2827872}. The dataset includes movie information (IDs, titles, genres, actors, release years) and user ratings (1 to 5). We use this dataset to (i) initialize user profiles, capturing their unique tastes and social traits, and (ii) train an SLM-powered user agent to accurately rate movies.

We conducted experiments on 200 users with 100-200 interactions each to simulate a user setting with long term interactions. Using GPT-4o, we distilled the historical interactions and clustered users into 4 personas with the KMeans++ algorithm. In order to create RAG data we retrieved top-1 interaction using cosine similarity. We fine-tuned the SLM using low-rank adapters of rank 256 (details in Appendix \ref{setup}). For fair comparison, the dataset for \(\theta_{\text{single}}\) was sub-sampled to match the largest dataset used for \(\theta_{\text{persona}}\). Moreover, given that each adapter in \(\theta_{\text{persona}}\) has access to different amounts of data, we trained two versions of \(\theta_{\text{single}}\): one using a dataset size equal to the smallest persona or \(\theta_{\text{single (min)}}\) and using the largest \(\theta_{\text{single (max)}}\). For each user, we merged their LoRA with the base SLM to predict movie ratings on a Likert scale of 1 to 5. We evaluated these agents using Root Mean Squared Error (RMSE), Mean Absolute Error (MAE), and Unrelated Response Rate (URR), which is the percentage of times the model produces an unrelated response.

\textbf{SLM vs LLM for Personalization}: We begin by comparing vanilla prompting on both \texttt{LLaMA-3} and \texttt{Phi-3-Mini}. We use both short-term and long-term memories as these models are not fine-tuned and not adding long-term memory could result in poor performance. In this particular instance, we observe in Table \ref{table:main}(No.1 and 4) that \texttt{LLaMA-3} beats \texttt{Phi-3-Mini} in terms of both MAE and RMSE. However, when we fine-tuned the SLM (No. 7), we observe that the performance increases significantly and beats \texttt{LLaMA-3} baseline. We also note that only using short term memory improves the frozen model's performance (No. 0 vs 1 and 3 vs 4). To understand the utility of user persona's we prompt \texttt{LLaMA-3} with only the long-term memories \(k=1\) and find that it achieves the least performance (No.2 vs 0 and 1). However, \texttt{LLaMA-3} does not produce an unrelated response at all while both non fine-tuned and fine-tuned versions of \texttt{Phi-3-Mini} produce unrelated responses across the runs.

\textbf{Does user clustering help?} We find that clustering users into specific personas increases performance when the model is fine-tuned on \(M_s\) + \(M_l\), as observed in Table \ref{table:main}. Specifically, in the comparison between No. 10 and Nos. 7 and 8, we see that No. 10 performs better, as the average RMSE of Nos. 7 and 8 is higher. Although No. 7 outperforms No. 10, it uses 2.5 times more training data, which accounts for its better performance. (As per Appendix B, Figure \ref{fig:rmse_mae}, the performance of \(\theta_{\text{persona}}\) improves with an increase in training dataset size.) Similarly, No. 9 achieves a lower RMSE compared to No. 5, but higher than No. 6. This discrepancy can be attributed to the lesser data diversity in each \(\theta_{\text{persona}}\)'s training set compared to \(\theta_{\text{single}}\), which could lead to performance drops among some persona clusters (Appendix B, Figure \ref{fig:words}.)

\textbf{Are long-term memories (\(M_l\)) useful?} In 2 out of 3 fine-tuning settings, we find that using long-term memories in addition to short-term memories provides better results than training only on short-term memories. This can be observed in the comparison between No. 5 and 7, where RMSE reduces by 0.181 and MAE by 0.056 Table (\ref{table:main}). Similarly, between No. 9 and 10, the latter achieves a 0.16 lower RMSE and 0.065 lower MAE. However, we observe an anti-pattern in the comparison between No. 6 and 8, where No. 6 achieves a lower RMSE, indicating that the model using long-term memory could not generalize well. We identify that our approach does not account for the quality of data retrieved during fine-tuning, and we leave this as a line of future work.

\begin{table}[t]
\centering
\label{table:performance_comparison}
\begin{tabular}{@{}lllllll@{}}
\toprule
\textbf{No.} & \textbf{Base Model} & \textbf{Adapter(s)} & \textbf{Memories Used} & \textbf{RMSE} ($\downarrow$) & \textbf{MAE} ($\downarrow$) & \textbf{URR} ($\downarrow$) \\
 \\ \midrule
0 & \texttt{LLaMA-3}   & -                           & \(M_s\) & \textbf{1.158}    & \textbf{0.847} & 0\%                          \\ \midrule
1 & \texttt{LLaMA-3}    & -                           & \(M_s + M_l\) & 1.221    & 0.883 & 0\%                          \\ \midrule
2 & \texttt{LLaMA-3}    & -                           & \(M_l\)       & 1.229              & 0.888               & 0\%  \\ \midrule[\heavyrulewidth]\midrule
3 & \texttt{Phi-3-Mini} & -                           & \(M_s\)       & \textbf{1.203}     & \textbf{0.863} & 4.23\%                         \\ \midrule
4 & \texttt{Phi-3-Mini} & -                           & \(M_s + M_l\) & 1.315     & 0.952 & \textbf{2.93\%}                         \\ \midrule
5 & \texttt{Phi-3-Mini} &  \(\theta_{\text{single (max)}}\) & \(M_s\)       & 1.312      & 0.940  & 1.30\%                          \\ \midrule
6 & \texttt{Phi-3-Mini} & \(\theta_{\text{single (min)}}\) & \(M_s\)       & 1.180      & 0.848  & 1.93\%                          \\ \midrule
7 & \texttt{Phi-3-Mini} & \(\theta_{\text{single (max)}}\)  & \(M_s + M_l\) & \textbf{1.150}    & \textbf{0.834} & \textbf{1.15\%}                       \\ \midrule
8 & \texttt{Phi-3-Mini} & \(\theta_{\text{single (min)}}\)  & \(M_s + M_l\) & 1.337    & 1.042 & 3.01\%                       \\ \midrule
9 & \texttt{Phi-3-Mini} & \(\theta_{\text{persona}}\) & \(M_s\)       & 1.292    & 0.937 & 2.29\%                       \\ \midrule
10 & \texttt{Phi-3-Mini} & \(\theta_{\text{persona}}\) & \(M_s + M_l\) & \uline{\textbf{1.171}}     & \uline{\textbf{0.881}} & \uline{\textbf{1.77\%}}        \\ \bottomrule
\end{tabular}
\vspace{0.6em}
\label{table:main}
\caption{Performance comparison of user agents built using different model and dataset strategies. The absence of an adapter means we did not fine-tune and only prompted the base model. \(\theta_{\text{single (min)}}\) signifies that a single LoRA was trained on \(\approx\) 2100 samples, while \(\theta_{\text{single (max)}}\) indicates it was trained on \(\approx\) 5200 samples. The presence of \(M_s\) or \(M_l\) indicates the type of memories that were used to train and evaluate the models. The numbers in bold show the best scores per category, and underlines show the comparison between single LoRA and persona LoRA.}
\end{table}
\vspace{-1em}

\section{Conclusions}
In conclusion, our work has demonstrated that distilling user interaction data into representative user knowledge, manifested as short-term and long-term memories, is highly effective in enabling user behavior agents. Furthermore, we have shown that fine-tuning SLMs with a low-rank adapter per persona significantly improves the model's ability to mimic user behavior, compared to training a single adapter for all users. In summary, our approach bridges the gap between offline evaluation metrics and real-world performance by utilizing scalable user simulation agents, especially in settings where users have extensive interaction histories. We are optimistic that our findings will pave the way for more scalable, personalized user interaction systems.

\section{Limitations and Future Work}



Our results demonstrate the promising use of SLMs over LLMs in user simulation settings. However, few limitations serve as future research directions. The distillation process, reliant on LLM prompting for quality outputs, can be slow with large datasets. Fine-tuning is constrained by computationally intensive hyperparameter tuning, and exploring other parameter-efficient methods beyond low-rank adapters could yield better results. While clustering users into personas shows positive outcomes, optimizing persona generation by considering more easily acquired user features is needed. We encourage further exploration of SLMs in user simulation for recommender systems and suggest training smaller models to generate enriched user personas using our knowledge distillation technique from LLMs.

\bibliographystyle{plain}
\bibliography{references}

\appendix

\section{Details on Models, Training and Dataset}

We conduct experiments on 200 users with 100-200 interactions per user. For the llm baselines and single LoRA runs, our training set size was 5132 rows, creating by sampling a fixed percentage of rows at random per user. The test set in all runs was kept to be the same, where we have 7496 rows. For persona setup, the training dataset size varied from 2k-5k for each cluster. In the persona setup, we also split the aforementioned test file per cluster based on the user id. Thus, each persona LoRA was used to evaluate only the users within that cluster. Also, both train and test sets have the same 200 users but different interactions, where interactions were split temporally following a 60:40 train test split. For the distillation of MovieLens dataset, we use GPT-4o to generate \( M_s \) and \( M_l \) by passing batches of interactions (batch size was kept to be less than or equal to 10). We rejected generations that were longer than 2000 words to ensure we are able to use the smaller context length provided by \texttt{Phi-3}. 

We cluster user profiles using KMeans++ to form \( K = 4 \) personas, determined by the Elbow method. To find the top-\( k \) nearest interactions, we use cosine similarity and append the enriched interaction text (from \( M_l \)) to the prompt and use \( k = 1\) for all experiments. During fine-tuning, we train low-rank adapters for key and query matrices in the attention layer with a rank of 256, alpha of 32, and dropout of 0.1. We use a learning rate of \( 3 \times 10^{-4} \) and train for 2 epochs using the AdamW optimizer. During inference, we prompt the model to generate 4 tokens, using a temperature of 0.3 and top-p set to 50 and look for the ground-truth using a regex search. To make comparison of \(\theta_{\text{single}}\) and \(\theta_{\text{persona}}\) fair, we subsample the dataset used to train dataset used to train \(\theta_{\text{single}}\) (by interactions per user) so the dataset size resembles the size of the largest dataset used to train \(\theta_{\text{persona}}\). We conduct all our experiments on 8xA100 GPUs with 80 GB of GPU memory on each core. We use the torchtune library for fine-tuning and evaluation, OpenAI APIs for distillation and sklearn metrics.

\label{setup}
\section{More on Persona}

\begin{figure*}[h]
    \centering
    \includegraphics[width=1\textwidth]{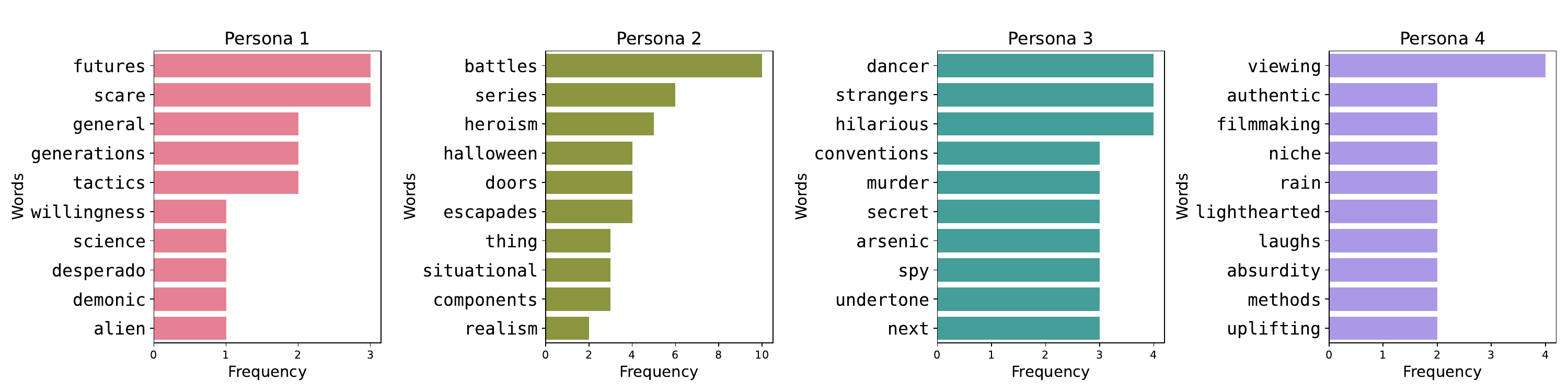}
    \vspace{-0.5cm}
    \caption{Top-10 distinct words found in each persona. We compute this by obtaining all nouns and adjectives using the Spacy \texttt{en\_core\_web\_sm} model, and finding one-vs-all set difference amongst the words in each cluster, retaining only top-10.}
    \label{fig:mainfig}
    \vspace{-0.2cm}
\label{fig:words}
\end{figure*}

In Figure \ref{fig:words}, we find that each persona cluster has a slightly different set of users, mainly by genre. For instance, persona 1 is all about method movies while persona 2 is about wars, battles and heroic stories. However, the token counts of these words are very low, indicating that these persona have overlaps as well.

\begin{figure}[h]
    \centering
    \begin{tikzpicture}
        \begin{axis}[
            xlabel={Train Dataset Size},
            ylabel={Value},
            legend style={at={(1.05,1)}, anchor=north west},
            grid=major,
            mark options={solid},
            width=10cm,
            height=6cm,
            xtick={2199, 3199, 3658, 5088},
            scaled ticks=false,
            tick label style={font=\small},
            label style={font=\small},
            nodes near coords,
            point meta=explicit symbolic
        ]
        \addplot[
            color=blue,
            mark=*,
            mark options={solid},
            thick
        ] table [meta index=2] {
            x y label
            5088 1.109585 1.109585
            3658 1.162975 1.162975
            3199 1.205178 1.205178
            2199 1.264979 1.264979
        };
        \addlegendentry{RMSE}

        \addplot[
            color=red,
            mark=square*,
            mark options={solid},
            thick
        ] table [meta index=2] {
            x y label
            5088 0.854753 0.854753
            3658 0.845188 0.845188
            3199 0.903695 0.903695
            2199 0.968322 0.968322
        };
        \addlegendentry{MAE}

        \end{axis}
    \end{tikzpicture}
    \caption{RMSE and MAE by Train Dataset Size}
    \label{fig:rmse_mae}
\end{figure}
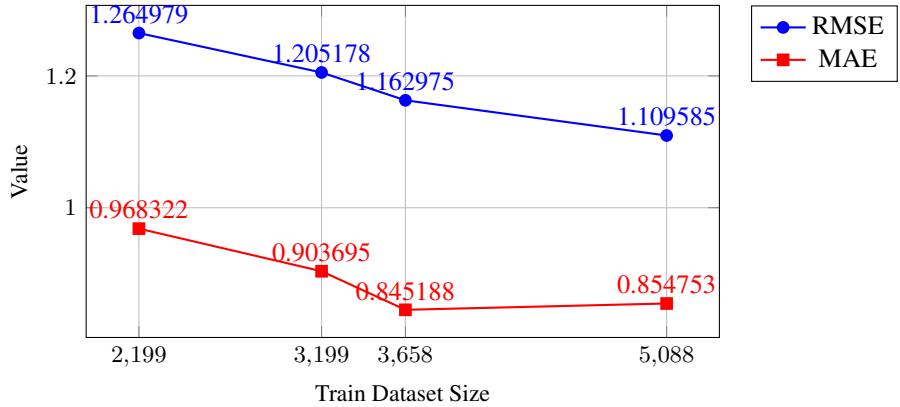

In Figure \ref{fig:rmse_mae}, we observe that the performance for each persona group depends on the size of the training set used. Both RMSE and MAE on the test set decrease (improve) as the training size increases. This indicates that larger training datasets contribute to better model performance, reducing both RMSE and MAE values. Each number show the performance of the fine-tuned model of the persona cluster on their respective test sets.

\thispagestyle{urls}

\newpage

\section{Data Samples}
\begin{customlisting}[Movie Rating prompt using Short Term Memory only]

PERSONA: You excel at role-playing. Picture yourself as a user exploring a movie recommendation system. You have the following social traits:. Your activity trait is described as: An Occasional Viewer, seldom attracted by movie recommendations. Only curious about watching movies that strictly align the taste. The movie-watching habits are not very infrequent. And you tend to exit the recommender system if you have a few unsatisfied memories.. Your conformity trait is described as: A Balanced Evaluator who considers both historical ratings and personal preferences when giving ratings to movies. Sometimes give ratings that are different from historical rating.. Your diversity trait is described as: A Cinematic Trailblazer, a relentless seeker of the unique and the obscure in the world of movies. The movie choices are so diverse and avant-garde that they defy categorization.. Beyond that, your profile description expressing your preferences and dislikes is: I have a strong appreciation for movies that blend wit, depth, and compelling narratives. My highest ratings are often reserved for films that offer a mix of dark humor, psychological intrigue, and intense drama. I am particularly drawn to complex characters and intricate plots, as seen in my preference for thrillers and dramas with strong performances by seasoned actors. I enjoy historical and action-packed films, especially those with a rebellious or heroic theme. My taste also includes a fondness for classic comedies and animated features that provide a nostalgic or heartwarming experience. I tend to favor movies with ensemble casts where each actor brings a unique element to the story. While I appreciate a variety of genres, I am less enthusiastic about films that lack depth or fail to engage me emotionally. Overall, my movie-watching habits reflect a desire for thought-provoking and emotionally resonant storytelling, with a particular interest in films that challenge conventional narratives and offer fresh perspectives. However, I have realized that I also appreciate visually stunning and action-packed crime thrillers, as well as heartwarming adventure films, even if they are in the children's genre. Additionally, I have a newfound appreciation for unique blends of genres, such as comedy and westerns, when they offer a fresh and engaging narrative. I also value films with strong character development and intricate plots, even if they are set in historical or war contexts, as long as they provide a compelling narrative and emotional depth. I have also come to appreciate older cinematic styles and classic films more than I initially thought, as long as they offer a compelling narrative and emotional depth. I also enjoy high-stakes, fast-paced narratives and unique storytelling approaches, as well as films with a strong visual and sci-fi element, provided they offer a compelling narrative..". The activity characteristic pertains to the frequency of your movie-watching habits. The conformity characteristic measures the degree to which your ratings are influenced by historical ratings. The diversity characteristic gauges your likelihood of watching movies that may not align with your usual taste.

TASK: Rate the movie given below on a likert scale from 1 to 5, where 1 means you hated it, 2 means you disliked it, 3 means you are neutral about it, 4 meaning you liked it and 5 means you absolutely loved it. Always respond with either one of 1, 2, 3, 4 or 5. Do not output anything else.

MOVIE: The movie Big Lebowski, The was released in the year 1998 is of Comedy, Crime, Mystery, Thriller genre with historical average rating of 3.74.
SUMMARY: A laid-back slacker gets caught up in a case of mistaken identity and embarks on a wild adventure involving bowling, kidnapping, and a rug that really tied the room together.. The cast of the movie is as follows: Jeff Bridges, John Goodman, Julianne Moore, Steve Buscemi, David Huddleston, Philip Seymour Hoffman, Tara Reid, Philip Moon, Mark Pellegrino, Peter Stormare, Flea, Torsten Voges, Jimmie Dale Gilmore, Jack Kehler, John Turturro

RATING: 
 
\end{customlisting}

\begin{customlisting}[Movie Rating prompt using both Short Term and Long Term Memories]


PERSONA: You excel at role-playing. Picture yourself as a user exploring a movie recommendation system. You have the following social traits:. Your activity trait is described as: An Occasional Viewer, seldom attracted by movie recommendations. Only curious about watching movies that strictly align the taste. The movie-watching habits are not very infrequent. And you tend to exit the recommender system if you have a few unsatisfied memories.. Your conformity trait is described as: A Balanced Evaluator who considers both historical ratings and personal preferences when giving ratings to movies. Sometimes give ratings that are different from historical rating.. Your diversity trait is described as: A Cinematic Trailblazer, a relentless seeker of the unique and the obscure in the world of movies. The movie choices are so diverse and avant-garde that they defy categorization.. Beyond that, your profile description expressing your preferences and dislikes is: I have a strong appreciation for movies that blend wit, depth, and compelling narratives. My highest ratings are often reserved for films that offer a mix of dark humor, psychological intrigue, and intense drama. I am particularly drawn to complex characters and intricate plots, as seen in my preference for thrillers and dramas with strong performances by seasoned actors. I enjoy historical and action-packed films, especially those with a rebellious or heroic theme. My taste also includes a fondness for classic comedies and animated features that provide a nostalgic or heartwarming experience. I tend to favor movies with ensemble casts where each actor brings a unique element to the story. While I appreciate a variety of genres, I am less enthusiastic about films that lack depth or fail to engage me emotionally. Overall, my movie-watching habits reflect a desire for thought-provoking and emotionally resonant storytelling, with a particular interest in films that challenge conventional narratives and offer fresh perspectives. However, I have realized that I also appreciate visually stunning and action-packed crime thrillers, as well as heartwarming adventure films, even if they are in the children's genre. Additionally, I have a newfound appreciation for unique blends of genres, such as comedy and westerns, when they offer a fresh and engaging narrative. I also value films with strong character development and intricate plots, even if they are set in historical or war contexts, as long as they provide a compelling narrative and emotional depth. I have also come to appreciate older cinematic styles and classic films more than I initially thought, as long as they offer a compelling narrative and emotional depth. I also enjoy high-stakes, fast-paced narratives and unique storytelling approaches, as well as films with a strong visual and sci-fi element, provided they offer a compelling narrative..". The activity characteristic pertains to the frequency of your movie-watching habits. The conformity characteristic measures the degree to which your ratings are influenced by historical ratings. The diversity characteristic gauges your likelihood of watching movies that may not align with your usual taste.

Given your persona and memory of some of movies you watched in the past, think carefully and rate the movie given in the end. Always use your memory at your own discretion as not everything is helpful. Also, historical average ratings of the movie mean an average rating of how other people have rated it, not you.

YOUR MEMORIES

MOVIE: The movie Falling Down was released in the year 1993 is of Action, Drama genre with historical average rating of 3.45.
MY MEMORY: A man's descent into madness and violence as he navigates through the frustrations of modern life.. The cast of the movie is as follows: Michael Douglas, Robert Duvall, Barbara Hershey, Rachel Ticotin, Tuesday Weld, Frederic Forrest, Lois Smith, Joey Singer, Ebbe Roe Smith, Michael Paul Chan, Raymond J. Barry, D.W. Moffett, Steve Park, Kimberly Scott, James Keane. I rated movie Falling Down (1993) as 4 because it offers a compelling narrative of a man's descent into madness and violence, aligning with my appreciation for psychological intrigue and intense drama. The strong performances by Michael Douglas and Robert Duvall, along with the thought-provoking and emotionally resonant storytelling, make it a highly engaging film.

TASK: Rate the movie given below on a likert scale from 1 to 5, where 1 means you hated it, 2 means you disliked it, 3 means you are neutral about it, 4 meaning you liked it and 5 means you absolutely loved it. Always respond with either one of 1, 2, 3, 4 or 5. Do not output anything else.

MOVIE: The movie Big Lebowski, The was released in the year 1998 is of Comedy, Crime, Mystery, Thriller genre with historical average rating of 3.74.
SUMMARY: A laid-back slacker gets caught up in a case of mistaken identity and embarks on a wild adventure involving bowling, kidnapping, and a rug that really tied the room together.. The cast of the movie is as follows: Jeff Bridges, John Goodman, Julianne Moore, Steve Buscemi, David Huddleston, Philip Seymour Hoffman, Tara Reid, Philip Moon, Mark Pellegrino, Peter Stormare, Flea, Torsten Voges, Jimmie Dale Gilmore, Jack Kehler, John Turturro

RATING:
\end{customlisting}
\end{document}